\documentclass[showpacs,10pt,twocolumn,prb]{revtex4-1}
\usepackage{amsmath}
\usepackage{amssymb}
\usepackage{graphics}
\usepackage{epsfig}

\begin{document}

\title{Synthesis, crystal structure and magnetism of $\beta $-Fe$%
_{1.00(2)}$Se$_{1.00(3)}$ single crystals}
\author{Rongwei Hu,$^{1,*}$ Hechang Lei,$^{1}$ Milinda Abeykoon,$^{1}$ Emil S.
Bozin,$^{1}$ Simon J. L. Billinge,$^{1,2}$ J. B. Warren,$^{3}$ Theo Siegrist,$%
^{4}$ and C. Petrovic$^{1,\ddag }$}

\affiliation{$^{1}$Condensed Matter Physics and Materials Science Department, Brookhaven
National Laboratory, Upton, NY 11973\\
$^{2}$Department of Applied Physics and Applied Mathematics, Columbia
University, New York, NY 10027, USA\\
$^{3}$Instrumentation Division, Brookhaven National Laboratory, Upton NY
1197, USA \\
$^{4}$Department of Chemical and Biochemical Engineering and National High
Magnetic Field Laboratory, Florida State University, Tallahassee, FL 32310,
USA}

\date{\today}

\begin{abstract}
Understanding iron based superconductors requires high quality impurity free
single crystals. So far they have been elusive for $\beta $-FeSe and
extraction of intrinsic materials properties has been compromised by several
magnetic impurity phases. Herein we report synchrotron - clean $\beta $-FeSe superconducting
single crystals grown via LiCl/CsCl flux method. Phase purity yields evidence for a defect induced weak ferromagnetism that
coexists with superconductivity below T$_{c}$. In contrast to Fe$_{1+y}$Te -
based superconductors, our results reveal that the interstitial Fe(2) site is not occupied
and that all contribution to density of states at the Fermi level must come
from in-plane Fe(1).
\end{abstract}

\maketitle

\section{Introduction}

The physics of complex superconductors, such as the cuprates and iron-based superconductors,
cannot be understood unless pure, high quality materials are available that
allow the intrinsic properties to be separated from extrinsic and impurity effects.
In FeAs and Fe(Se)Te,
just as in high-T$_{c}$ cuprate and heavy fermion materials,
competing or coexisting magnetic order is closely associated with
superconductivity.~\cite{Mazin1} This suggests proximity to a
magnetic critical point and an unconventional origin of superconductivity where
spin fluctuations may contribute to pairing.~\cite{Mazin3,Kuroki,Drew} The
observation of weakly localized rather than itinerant magnetism sensitive to
structural changes raises the fundamental question of how strongly correlated
are the charges in Fe superconductors and what is the origin of the magnetic
order?~\cite{Cvetkovic,YangW,Bao1} Of particular interest is superconducting
$\beta $-FeSe, a compensated semimetal, without a crystallographic charge
reservoir, that superconducts at about 8~K without any carrier doping.\cite%
{Hsu} It has a giant pressure coefficient of $T_c$ of 9.1~K/GPa
enhancing T$_{c}$ up to a maximum of 37~K, the third highest known
critical temperature for any binary compound.\cite{Medvedev}

A major obstacle in understanding intrinsic magnetism in $\beta $-FeSe has been the purity of
the material itself. Magnetic impurities such as $\alpha $-FeSe, Fe$_{7}$Se$_{8}$, Fe%
$_{3}$O$_{4}$ and elemental Fe are ubiquitous in all as-grown crystals and sometimes
polycrystals.\cite{Zhang,Patel,McQueen,Phelan} They contribute to the
large ferromagnetic (FM) background, seen in the M-H loops below
superconducting T$_{c}$. Modification of the original Fe-Se
phase diagram near 1:1 stoichiometry suggested that $\beta $-FeSe is not
stable at the room temperature since it converts to hexagonal $\alpha $-FeSe
below 300$^{\circ}$C.\cite{McQueen,Okamoto} Consequently, the absence of an exposed liquidus surface in
the binary alloy phase diagram and the metastable nature of the superconducting FeSe
are considered to be prohibitive and insurmountable factors for single
crystal preparation using standard synthesis methods.

Here we describe a synthetic approach that yields stoichiometric
and phase pure material and we report intrinsic structural and magnetic properties of
superconducting $\beta $-FeSe. These include evidence for defect - induced weak ferromagnetism (WFM) and the absence of interstitial Fe(2) whose occupancy governs the magnetic
and structural phase diagram in isostructural Fe$_{1+y}$Te.\cite{Bao1}

\section{Experiment}

Powders of LiCl
and CsCl, elemental Fe and Se were added together with the flux into an alumina crucible and sealed under partial Ar
atmosphere. The ampoule was heated to a homogenization temperature of 715$^{\circ}$C, where it
was kept for 1 h and then removed into a preheated furnace at
457$^{\circ}$C. After slow cooling to 300$^{\circ}$C, it was quenched in water.

Medium resolution, room temperature (300 K) X-ray diffraction measurements
were carried out at X7B beam-line at National Synchrotron Light Source (NSLS)
at the Brookhaven National Laboratory, using a 0.5~mm$^2$ monochromatic beam of 38.92 keV ($\lambda$ = 0.3184 \AA).
Pulverized sample was filled into a 1 mm diameter cylindrical Kapton capillary and the data
collection was carried out in a forward scattering geometry using a Perkin Elmer 2-D detector
mounted orthogonal to the beam path 378.3 mm away from the sample.

Single crystals of $\beta$-FeSe were also mounted on glass fibers for examination
using an Oxford - Diffraction Xcalibur 2 CCD 4-circle diffractometer with
graphite monochromated MoK$\alpha $ radiation. Elemental and microstructure analysis were performed on several $\beta $-FeSe crystals as
well as on the particular crystals chosen for resistivity and magnetization
using energy-dispersive X-ray spectroscopy in a JEOL JSM-6500 scanning
electron microscope (SEM).

Sample dimensions were measured with an optical microscope Nikon SMZ-800
with 10 $\mu $m resolution and M/H values were corrected for straw
background at each (T,H) of the measurement, real sample volume and
demagnetization factor. Thin Pt wires were attached to electrical contacts
made of Epotek H20E silver epoxy for a standard four-probe measurement with
current flowing in the (101) plane of the tetragonal structure. Magnetization and
resistivity measurements were carried out in a Quantum Design MPMS and PPMS
respectively.

\section{Results}

Fig. 1 shows Fe-Se and LiCl-CsCl phase diagrams.\cite{Okamoto,Chartrand}
The superconducting PbO-type $\beta $-FeSe is a low temperature
crystallographic phase that decomposes into Fe and hexagonal NiAs phase (${\alpha}$-FeSe) at 457$^{\circ}$C (Fig. 1(a)).
It coexists with hexagonal ${\alpha}$-Fe$_{7}$Se$_{8}$ below 300$^{\circ}$C for certain Fe-Se stoichiometry.\cite{Okamoto} Previous attempts (for example Refs. 10, 11, 16) to
prepare single crystals of $\beta $-FeSe involved nucleation and growth using KCl/NaCl flux or vapor transport reactions.
We choose a LiCl - CsCl flux method of synthesis due to the presence of a
low temperature eutectic at 326$^{\circ}$C, well
below the decomposition temperature of $\beta $-FeSe.\cite{Chartrand} The crystal growth
possibly includes nucleation of Fe$_{7}$Se$_{8}$ above 700$^{\circ}$C
and structural phase transition at low temperatures. As opposed to crystals grown in KCl,\cite{Mok} the
low temperature eutectic (Fig. 1(b)) allows for complete
transition to tetragonal $\beta $-FeSe from (457 - 300)$^{\circ}$C in a large
fraction of crystals grown in a batch. Platelike FeSe crystals with the (101) plane exposed
and elongated in one direction up to 1.5 $\times $ 0.5 $\times $ 0.05 mm$%
^{3} $ can be separated by dissolving the flux in de-ionized water and
rinsing in ethanol.

\begin{figure}[tbp]
\centerline{\includegraphics[scale=1.2]{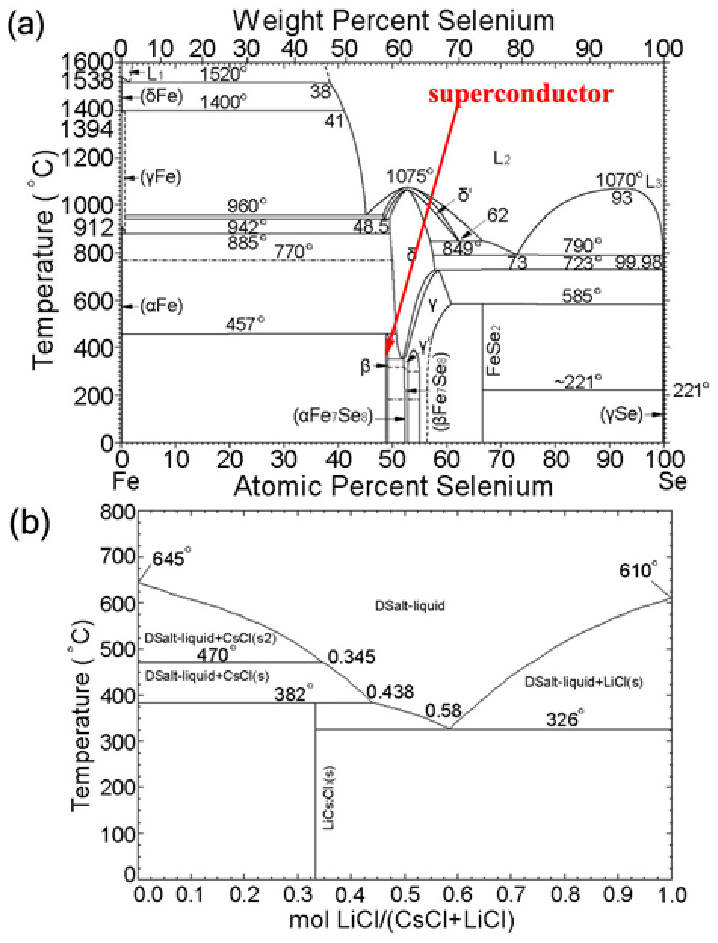}} \vspace*{-0.3cm}
\caption{(Color online) Fe-Se (a) and CsCl - LiCl phase diagrams (b). The presence of the low temperature eutectics (b) enables
long annealing in the liquid below 457$^{\circ}$C and complete transition from ${\alpha}$-FeSe to ${\beta}$-FeSe in the large fraction of the crystals in a batch.}
\end{figure}

\begin{figure}[tbp]
\centerline{\includegraphics[scale=1.0]{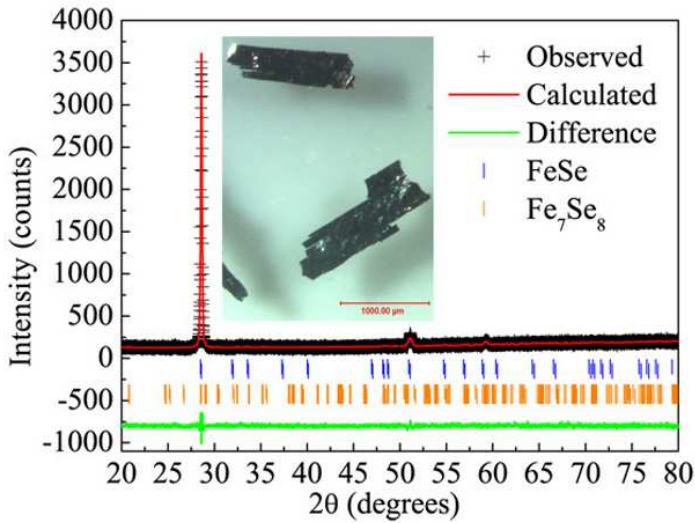}} \vspace*{-0.3cm}
\caption{(Color online) Powder X-ray diffraction spectra on a $\protect\beta$-FeSe single crystal
shows no impurity phases and (h0l) crystal orientation at 300 K. The data were shown by (+), the fit is given by the top solid line and the difference curve (bottom solid line) is offset for clarity. Allowed crystallographic reflections are given as vertical tick marks for $\beta $-FeSe (top line) and Fe$_{7}$Se$_{8}$ (bottom line). Bragg peak (201) for 2$\protect\theta$ = 51.1 clearly distinguishes $\beta $-FeSe from Fe$_{7}$Se$_{8}$, in addition to magnetic properties. Inset shows typical $\beta $-FeSe single crystals.}
\end{figure}

\begin{figure}[tbp]
\centerline{\includegraphics[scale=0.6]{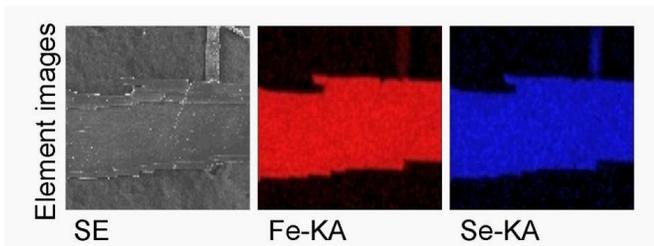}} \vspace*{-0.3cm}
\caption{(Color online) Microprobe electron density maps of as-grown $\protect\beta $-FeSe
single crystals.}
\end{figure}

\begin{figure}[tbp]
\centerline{\includegraphics[scale=0.7]{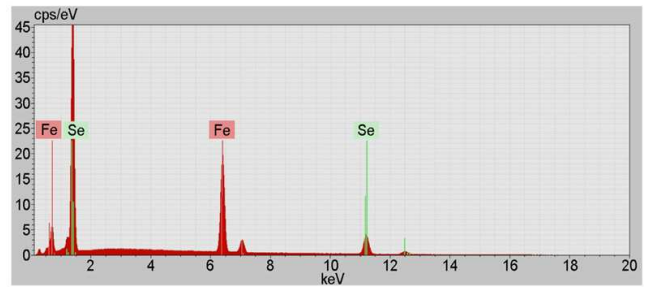}} \vspace*{-0.3cm}
\caption{(Color online) Elemental analysis of as-grown $\protect\beta $-FeSe single crystals.}
\end{figure}

\begin{figure}[tbp]
\centerline{\includegraphics[scale=0.9]{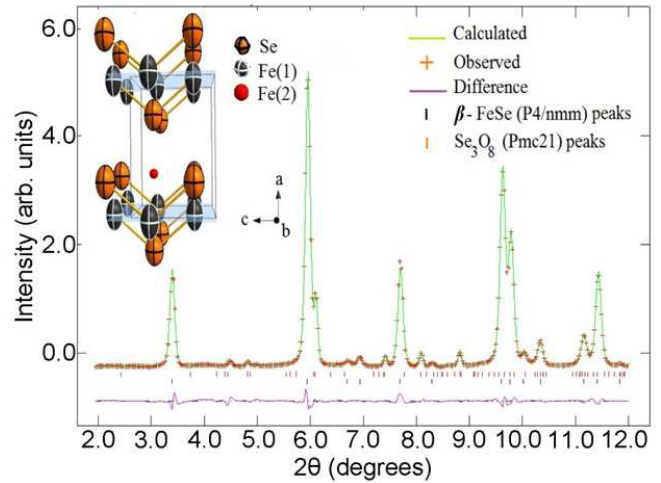}} \vspace*{-0.3cm}
\caption{(Color online) Structural refinement of FeSe synchrotron powder X-ray diffraction data over a narrow range of scattering angle, 2$\theta$, taken at 300 K. Background subtracted data are shown as (+), fit is given as a top solid line, and the difference curve (bottom solid line) is offset for clarity. Allowed crystallographic reflections are given as vertical tick marks: $\beta $-FeSe phase (bottom), and Se$_{3}$O$_{8}$ impurity phase (top) due to sample oxidation. Inset shows $\beta$-FeSe structure.}
\end{figure}

Crystals of $\beta $-FeSe were separated from $\alpha $-Fe$_{7}$Se$_{8}$ by a
combination of a permanent
magnet and a powder X - ray diffraction (XRD) spectra from a Rigaku Miniflex with CuK$\alpha $ radiation ($\lambda $ =
1.5418 \AA ). We observed only (h0l) peaks (Fig. 2) of the tetragonal phase in
selected crystals for further analysis (Fig. 2 inset). Crystals contaminated with
Fe$_{7}$Se$_{8}$ and/or oxides showed additional peaks in the spectra.

Fig. 3 and Fig. 4 show electron density maps and elemental
analysis of as-grown $\beta $-FeSe single crystals. The relative stoichiometry of multiple points on
the as-grown crystals were measured and a composition of Fe$_{0.99(4)}$Se
was obtained. Electron density maps of these crystals confirmed a uniform
distribution of Fe and Se.

Crystals are moderately air sensitive. After one week of
air exposure, a selenium oxide layer is visible on the crystal surface. After about
a month of air exposure Fe$_{3}$O$_{4}$ is detectable in laboratory X-ray diffraction (XRD) and
by the observation of a Verwey transition in M(T).\cite{Yang} The first attempts to carry out
synchrotron XRD experiments on samples that were exposed to air for
several months revealed the presence of multiple additional phases including
appreciable amounts of several selenium oxide and iron oxide phases. It has
been found that while pure stoichiometric FeSe crystals can be grown, these
tend to degrade through oxidation over the course of time. This suggests
that the surface of FeSe crystals may be Se-terminated and that selenium
oxide forms first, with further degradation involving iron oxide phases or Fe%
$_{2}$[SeO$_{3}$]$_{2}$O as well. Results reported here are based on samples
whose exposure to air was minimized, and only traces of selenium oxide were
found in crystals that were pulverized for the powder XRD
experiments.

Synchrotron powder XRD data of FeSe sample were successfully
refined using a two phase structural model (Fig. 5). The best
fit contained 96.1 mol\% (90.7 wt\%) of FeSe of \textit{\ P4/nmm} space
group, with a = 3.7622(2) \AA , c = 5.5018(5) \AA , with Se at
(1/4,1/4,0.2624(1)) and Fe at (3/4,1/4,0). Compared to the high T$_{c}$ stoichiometric polycrystalline $\beta $-FeSe,\cite{McQueen,LiZ} the  unit cell parameters are reduced by 0.3\% (a -axis) or 0.4\% (c-axis), whereas c/a is smaller or identical. The anisotropic atomic displacement parameters (ADP) ratio U$_{33}$/U$_{11}$ is the ratio of thermal vibrations along crystallographic c and a axes in tetragonal strucutre. The U$_{33}$/U$_{11}$ was 1.12 for Se and 1.41 for Fe. To illustrate this,
anisotropic ADP's are shown as exaggerated thermal ellipsoids in Fig. 5 inset.
In Van der Waals bonded crystals, such as FeSe, U$_{33}$/U$_{11}$ ADP ratio is expected to
be larger than 1 and the observed ratios are within the expected range. A somewhat
larger ADP ratio of Fe, 1.41, suggests that it is underbonded and can move along
the c-axis. The FeSe$_{4}$ units are found to deviate from perfect tetrahedra, with an
Fe-Se distance of 2.379(5)  \AA , and tetrahedral angles of 104.5(5) and 112.0(5) degrees. The anisotropic ADP ratio (tetrahedral angles) are smaller (equal) than the values obtained for $\beta $-FeSe polycrystals on powders containing several magnetic impurity phases.\cite{McQueen,Margadonna,Margadonna2}However, the anisotropic ADP ratio observed here is similar to values found in pure Fe$_{1.08}$Te.\cite{Onoda} As expected, the observed tetrahedral bond angles deviate from the ideal tetrahedral angle found in iron based superconductors with optimal T$_{c}$.\cite{Kimber} The second phase, constituting 3.9 mol\% (9.3 wt\%), was found to be Se$_{3}$O$_{8}$ with
\textit{Pmc21} space group with the refined lattice parameters a = 4.977(1) \AA , b = 4.388(2) \AA , c =
15.377(2) \AA . No other phases were observed.
Within the main phase we investigated in detail the issue of stoichiometry and
occupancy of the interstitial site, Fe(2) at (3/4,3/4,z) (Fig. 5 inset). The stoichiometry
was found to be Fe$_{1.00(2)}$Se$_{1.00(3)}$.

Difference Fourier analysis (DFA) is a standard method to find missing electron densities
in refined atomic structures. In this technique, the difference between the observed and the calculated
(model based) Fourier maps is used to locate missing atoms in atomic structures.
In this study, DFA did not reveal any appreciable electron density
at the interstitial Fe(2) positions. Attempts to explicitly refine Fe(2) site occupancy yielded 0.00(1),
in agreement with the DFA and strongly suggesting that no iron resides on this site. However, DFA
indicated a possibility for additional electron density in the vicinity of Se. This, along with the observation
of relatively large anisotropic ADPs (U$_{33}$) of Se and Fe, may point to the presence of static and/or dynamic
disorder associated with these sites. A small number of Se vacancies
may lead to relaxation of the surrounding Fe atoms, resulting in static and/or dynamic disorder.

The resistivity $\rho $(T) of LiCl/CsCl flux grown crystals smoothly changes to linear at low temperature where the onset of T$_{c}$ and zero resistivity were observed at temperatures
about 1 K or more higher than in polycrystals (Fig. 6).\cite{Hsu} The residual resistivity ratio (RRR) of 14 indicates good crystal quality whereas the single crystal diffraction pattern shows no impurities present. Reciprocal space planes (hk0)
and (h0l) were reconstructed from several series of CCD frames (inset in Fig. 6). A mosaic structure is observed
perpendicular to the c - axis, consistent with the arrangement
of FeSe layers in the structure. The observed X-ray reflections are all
consistent with the $\beta $-FeSe structure. M/H exhibits weak temperature dependence for
both H $\perp $ (101) and H$\parallel $(101) (Fig. 7(a)). Below 135 K the M/H signal drops and then
remains constant below the structural transition temperature $\sim100$ K.\cite
{Hsu} This is more pronounced for H $\perp $ (101). Low temperature M/H taken in
H = 10 Oe confirms superconductivity (Fig. 7(b)) below 9.0(2) K.
Extrapolation of 4$\pi \chi $ data to T = 0 gives about 60 \% of diamagnetic
screening. Complete $\rho $ transition and partial superconducting volume
fraction have been observed in SmFeAsO$_{1-x}$F$_{x}$ and CaFe$_{1-x}$Co$_{x}$%
AsF,\cite{Drew,Takeshita} where temperature dependent magnetic moment
coexists and inversely scales with the superconducting volume fraction.

\begin{figure}[tbp]
\centerline{\includegraphics[scale=0.75]{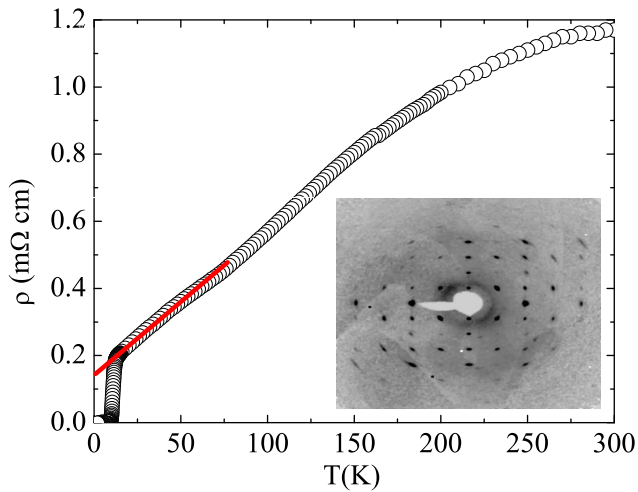}} \vspace*{-0.3cm}
\caption{(Color online) Temperature dependence of the resistivity for current in (101)
plane with T$_{c}$ onset and zero resistance of 12.0(1) and 9.2 (2) K, respectively.
Inset shows precession pattern of the (101) plane of
the same single crystal. All spots can be indexed within $\protect\beta $%
-FeSe space group with no impurities present. The large mosaic is visible
but the impurity free unit cell parameters are in agreement with published
(see text).}
\end{figure}

\begin{figure}[tbp]
\centerline{\includegraphics[scale=0.7]{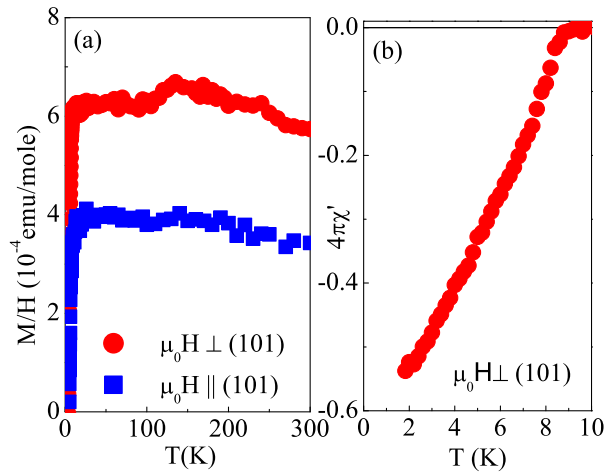}} \vspace*{-0.3cm}
\caption{(Color online) Anisotropy in high temperature M/H for H = 10 kOe
and superconducting volume fraction for H $\perp $ (101) measured in 10 Oe field.}
\end{figure}

In $\beta $-Fe$_{1.01}$Se polycrystals contaminated with magnetic impurity
phases a static moment was found above 1 GPa, \cite{Bendele}
was ascribed to traces of Fe impurity at ambient pressure due to its weak
nature,\cite{Khasanov,Fedorchenko} or was not detected. \cite{Medvedev, Imai} In our crystals,
s - shape of M(H) for H$\parallel $(101) (Fig. 8(a) is typical of a type-II
superconductor with a superimposed isotropic weak ferromagnetic (WFM) moment both above and below T$%
_{c}$.\cite{Patel} In contrast, the M(H) curves are symmetric for H $\perp $(101) at T =
1.8 K (Fig. 8(b) within the experimental resolution (0.05 emu/cm$^{3}$) with
no evidence for WFM below T$_{c}$.

\begin{figure}[tbp]
\centerline{\includegraphics[scale=1.2]{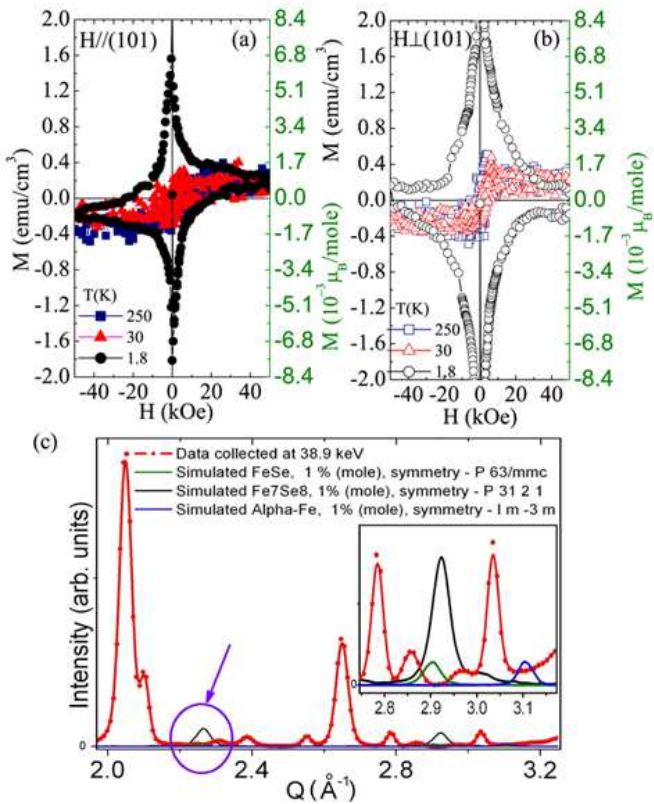}} \vspace*{-0.3cm}
\caption{(Color online) Magnetization isotherms below and above T$_{c}$ for magnetic field parallel (H$\parallel $(101)) (a)
and perpendicular to (101) plane (H $\perp $ (101)) (b). Right hand side scale shows magnetic moment assuming molar mass of $\beta $-FeSe (c) Comparison of background -- subtracted diffraction data with simulated impurity
phases at 300 K. Selected 2$\theta
$ region is shown where simulated impurity peaks are clearly visible.}
\end{figure}

\section{Discussion}

Is the observed WFM intrinsic or extrinsic? Fig. 8(c) shows background-subtracted data compared to
a simulated pattern for 1 mole~\% of commonly found magnetic phases. Based on the scattering power and distribution of peaks in our synchrotron
powder XRD we can exclude contamination by $\alpha $-FeSe and Fe$_{7}$Se$_{8}$ (Fig. 8(c) and Appendix A), leaving only elemental Fe, or some
unknown ferromagnetic phase with lattice periodicity commensurate with $\beta$-FeSe,~\cite{Mang} as a possible source of magnetic contamination. By focusing on information sensitive to the total number of Fe ions in
a given volume we make several observations that support the intrinsic WFM
scenario in as-grown crystals. M\"{o}ssbauer quadrupole splittings and isomer shifts for
binary Fe-Se materials differ by up to a factor of 4 whereas the signal component originating in the $\beta $-FeSe
phase dominates the M\"{o}ssbauer spectrum below room temperature in 50 kOe even for the sample containing
more than 6 mole \% of $\alpha $-FeSe, Fe$_{3}$O$_{4}$ and elemental Fe impurities.\cite%
{McQueen,KimE,Blachowski} Moreover, the coercive field of bulk Fe or Fe nanoparticles ($\mu _{0}$H$_{c}$ $\leq $ (0.4 - 2.5)
kOe)is several orders of magnitude lower than in Fig. 8(a,b) (15 kOe).\cite{Brown,Carvell}
Furthermore we observe the signature of the structural phase transition in the normal state (Fig. 7(a)), implying that a considerable fraction of the M(T) signal must come from the $\beta $-FeSe. Finally, WFM due to an unknown Fe - based high T$_{c}$ ferromagnetic phase\cite{Mang} is unlikely to have $\mu _{0}$H$_{c}$ in kOe range and
is expected to provide an isotropic constant (or increasing) background (bias) to M(H) loops \textit{both} above and below T$_{c}$, as in ref. 11.
This is in contrast to anisotropic M(H) below T$_{c}$ (Fig 8(a,b)).
Note that M(H) in Fig. 8(b) is symmetric with respect to the M = 0 line. This suggests that most of WFM signal in
$\beta $-FeSe crystals is unrelated to extrinsic impurities. The WFM signal is well reproduced in several crystals grown in one batch and in crystals grown from multiple batches, whereas the magnitude of WFM increases in time (Appendix B).

The proposed spin-density-wave (SDW) mechanism of magnetic order may not
apply to iron chalcogenides and perhaps it could be more complex even for
iron pnictides.\cite{Balatsky,Yin} Density functional calculations
indicate that magnetic state with 0.15 $\mu _{B}$/mole is induced in Fe$_{1-x}$Se for $x$ = 0.0625.\cite{Li2}
The saturation moment $\left\vert M_{s}\right\vert $ observed (Fig 8(a,b)) is about 1/150 smaller and
would correspond to an Fe deficiency
of less than 0.4 atomic~\%, a value below resolution limit of our diffraction measurement. However, in this calculation \cite{Li2} the partial density of states (PDOS) of
Fe(2) dominates the total density of states (DOS) and, more importantly,
the stoichiometry variation. Negligible occupancy of Fe(2) within experimental
error in our crystals implies that Fe nonstoichiometry is not the dominant
mechanism of WFM. On the other hand, it is possible that some fraction of
the WFM arises due to a Se vacancy induced magnetic cluster.\cite{Lee} The Se --
Se distances are Van der Waals distances and may produce Se -- Se time
dependent bonding. We cannot distinguish between static and dynamic
displacements but since the refinement results are giving 1:1 stoichiometry,
then vacancies could be equally distributed on both sites. In
particular, theory predicts that the main effect of Se displacement would be
to shift Fe(1) towards the vacant site, shifting the Fermi level E$_{F}$
into a sharp peak in the DOS that would promote a more stable magnetic state
than in material without Se defects.\cite{Lee} The net moment at 1.25 mole
\% Se deficiency is expected to be in 10$^{-2}$ $\mu _{B}$/mole range. This
is arising from both Fe(1) and Fe(2) contributions. Since the theoretical
contribution of Fe(2) PDOS at the Fermi level is about 50\% of the total DOS,\cite{Lee} the calculated moment is somewhat higher but generally in line with
the observed $\left\vert M_{s}\right\vert $ $\sim $ (1.0$\pm $0.5)$\cdot $10$%
^{-3}$ $\mu _{B}$/mole above T$_{c}$ in our crystals with neglibible Fe(2)
occupancy. This is different from most FeAs
superconductors where small moment magnetic order from a SDW mechanism is found below the
structural transition. This is also different from Fe$_{1+y}$Te where subtle
crystal chemical effects, with both Fe(1) and Fe(2) occupied, induces WFM and
structural and magnetic differences below the magnetostructural transition
at 75 K - 55 K.\cite{Bao1} Since lattice distortions were also recently
found\cite{Saha} to induce both superconducting and magnetic phases in SrFe$%
_{2}$As$_{2}$, this suggests that nanoscale defects and short range
structural features are important in a wider class of iron based
superconductors. Indeed, there is emerging evidence that both conducting and
magnetic properties in the recently discovered K$_{x}$Fe$_{2-y}$Se$_{2}$ superconductors are
governed by Fe vacancies. \cite{Bao2,Bao3} In $\beta $-FeSe, defect - induced magnetism
coexists with superconductivity that sets in far below T$_{c}$. Though rather unlikely, our analysis allows for some contribution of different impurity phase to WFM. The unknown high temperature FM phase would be present in quantities too small to detect by diffraction and/or would have the lattice periodicity commensurate with $\beta $-FeSe and its moment would anisotropically diminish between 30 K and 1.8 K.

In summary, single phase
superconducting single crystals of $\beta $-FeSe have been synthesized. Unlike isostructural Fe$_{1+y}$Te, the Fe(2) site is not occupied at all in these samples. The ADP anisotropy is consistent with dynamic disorder/defects associated with Fe and Se sites and/or Se vacancies. We present evidence for intrinsic defect induced WFM which anisotropically diminishes with an increase in the superconducting volume fraction.

\section{Acknowledgments}

We thank Sang-Wook Cheong and Hai-Hu Wen  for useful discussion and
J. C. Hanson for help with XRD measurements. This work was carried
out at the Brookhaven National Laboratory, which is operated for the U.S.
Department of Energy by Brookhaven Science Associates DE-Ac02-98CH10886.
This work was in part supported by the U.S. Department of Energy, Office of
Science, Office of Basic Energy Sciences as part of the Energy Frontier
Research Center (EFRC), Center for Emergent Superconductivity (CES). A
portion of this work was performed at the National High Magnetic Field
Laboratory, which is supported by NSF Cooperative Agreement No. DMR-0654118,
by the state of Florida, and by the DOE.

\section{Appendix A: Magnetic impurities from diffraction}

Magnetization measurements were performed on a sample volume V = 4.532$\cdot $10$^{-5}$ cm$^{3}$
(a = 0.223476 cm, b = 0.0555613 cm, c = 0.00365025 cm), corresponding to
(1.45$\pm$0.23)$\cdot $10$^{-5}$ emu MPMS signal. Assuming that magnetization (M)
of $\beta $-FeSe is negligible when compared to impurity magnetization, we discuss
possible magnetic impurity levels.

First we assume that the sample contains 1 mole \% of elemental $\alpha $-Fe since any higher Fe content would have been detected (Fig. 8(c)). In order
to obtain volume ratio we need to multiply mole (i.e. formula unit) ratios
of Fe and $\beta$-FeSe by (M/D) where M is the molar mass and D is the
density in g/cm$^{3}$. Using M(Fe) = 55.8 g/mole, D(Fe) = 7.83 g/cm$^{3}$, M(%
$\beta $-FeSe) = 134.8 g/mole, D($\beta $-FeSe) = 5.72 g/mole, we get 0.3
volume \%.

\begin{equation}
\frac{1mole(Fe)\cdot 7.12\frac{cm^{3}}{mole}}{100mole(\beta -FeSe)\cdot 23.56%
\frac{cm^{3}}{mole}}=0.003
\end{equation}

That would correspond to 1.37$\cdot $10$^{-7}$ cm$^{3}$ volume of Fe in our
sample. What would be its magnetic contribution? By dividing D/M we get 0.14
mole/cm$^{3}$ of Fe. In 1.37$\cdot $10$^{-7}$ cm$^{3}$ iron volume we have
1.92$\cdot $10$^{-8}$ mole Fe. Iron saturation moment is 2.2 $\mu _{B}$/mole, therefore that volume would have M = 4.22$\cdot $10$^{-8}$ $%
\mu _{B}$. Using conversion factor 5585 emu/$\mu _{B}$ we obtain M = 2.36$%
\cdot $10$^{-4}$ emu. This is more than 100 \% of raw M (emu) signal in
MPMS. Hence, it is possible that Fe impurity content in our crystal (mole \%
and consequently volume \%) is below detection capacity of synchrotron
powder X-ray diffraction data since only a fraction of 1 mole \% Fe would
generate such signal. This is about 0.05 atomic \% of Fe.

Similarly, 1 mole \% impurity of $\alpha $-FeSe is approximately identical
to 1 \% volume since $\alpha $-FeSe has the identical molar mass M and 95\%
of $\beta $-FeSe density. Therefore 1 \% mole $\alpha $-FeSe impurity would
correspond to 4.5$\cdot $10$^{-7}$ cm$^{3}$ volume (using eq. (1)). Applying
the same argument as above, 4.5$\cdot $10$^{-7}$ cm$^{3}$ volume $\alpha $%
-FeSe contains 1.9$\cdot $10$^{-8}$ mole $\alpha $-FeSe. The $\alpha $-FeSe
saturation moment is M$_{s}$ $\sim $ 0.2 $\mu _{B}$/mole.\cite{Hirone}
Therefore such volume would contribute M = 3.81$\cdot $10$^{-9}$ $\mu _{B}$.
Using conversion factor 5585 emu/$\mu _{B}$ we obtain M = 2.13$\cdot $10$%
^{-5}$ emu. Our raw MPMS signal is 70 \% that value. However, 0.7 mole \% of $\alpha $-FeSe would have been detected, if present (70
\% of its peak height, (Fig.8(c), main text)).

Finally, 1 mole \% of Fe$_{7}$Se$_{8}$ impurity (using M(Fe$_{7}$Se$_{8}$) =
1022.59 g/mole, D(Fe$_{7}$Se$_{8}$) = 6.43 g/cm$^{3}$ and eq. (1)) would
correspond to 6.75 \% of measured sample volume, which is 3.059$\cdot $10$%
^{-6}$ cm$^{3}$. Since the expected Fe$_{7}$Se$_{8}$ saturation moment is M$%
_{s}$ $\sim $ 80 (emu/cm$^{3}$)\cite{Kamimura} we would expect that such
volume would contribute with 2.44$\cdot $10$^{-4}$ $\mu _{B}$. Our raw MPMS
emu signal is 6 \% of that value, but still above the threshold of
scattering power detection in synchrotron experiment. If we
multiply the observed intensity of 2.92 \AA $^{-1}$ Fe$_{7}$Se$_{8}$ peak
(Fig. 8(c) main text) by 0.06, it is still above the background.

\section{Appendix B: Reproducibility, impurities and magnetic signal over time}

Figure 9 shows Rietveld refinement over the full 2$\theta $ range with only selenium oxide present due to oxidation.

\begin{figure}[tbp]
\centerline{\includegraphics[scale=0.9]{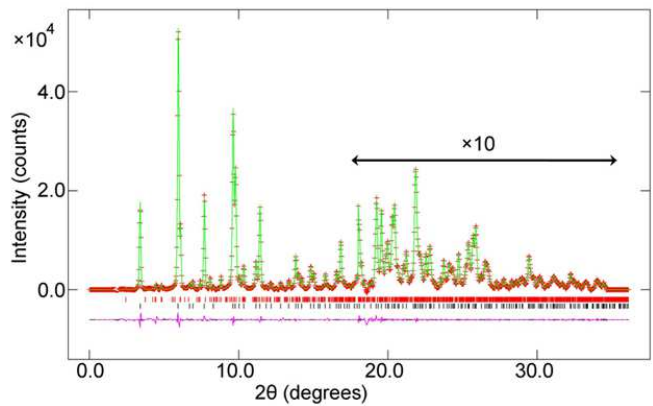}} \vspace*{-0.3cm}
\caption{(Color online) Structural refinement of FeSe synchrotron powder X-ray diffraction
data over the full 2$\protect\theta $ range taken at 300 K. Data (background subtracted)
are shown as (+), fit is given as a top solid line, and the
difference curve (bottom solid line) is offset for clarity. Bottom vertical tick marks
represent reflections in the main $\protect\beta $-FeSe phase (\textit{P4/nmm%
}), while top tick marks denote reflections in Se$_{3}$O$_{8}$ (%
\textit{Pmc21}).}
\end{figure}

\begin{figure}[tbp]
\centerline{\includegraphics[scale=0.75]{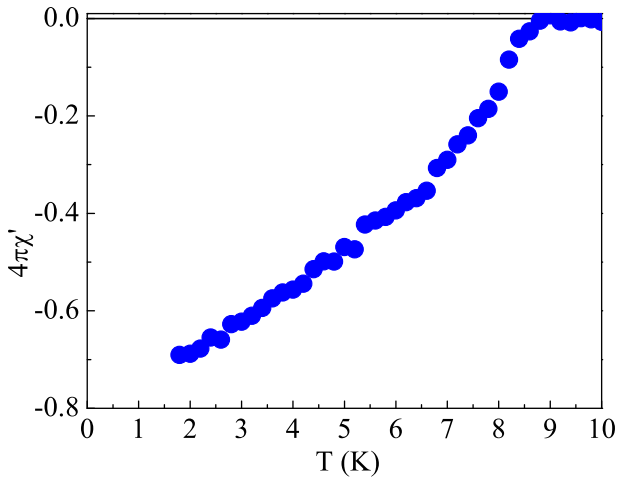}} \vspace*{-0.3cm}
\caption{(Color online) An example of superconducting T$_{c}$ of independently grown crystal taken in 10 Oe field.}
\end{figure}

\begin{figure}[tbp]
\centerline{\includegraphics[scale=0.7]{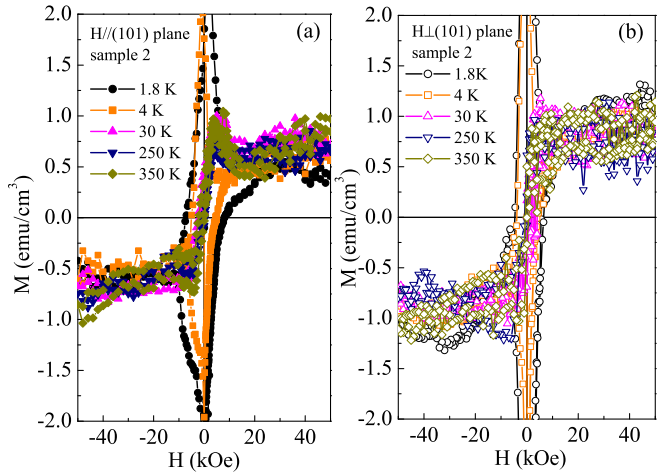}} \vspace*{-0.3cm}
\caption{(Color online) (a) Sample 2 M/H above and below T$_{c}$ for H$\parallel $(101) near M =
0. (b) Sample 2 M/H above and below T$_{c}$ for H$\perp $(101) near M = 0.}
\end{figure}
Powder X-ray diffraction taken on single crystals found no extrinsic peaks
in 10 out of 10 crystals that were separated by a magnet. Only
clean $\beta $-FeSe single crystals are used in further analysis. Samples
that were contaminated by Fe$_{7}$Se$_{8}$ (from synthesis or oxidation)
showed dominant FM hysteresis loop below T$_{c}$ with only traces of type-II
superconductor MvsH. Resistivity and magnetization data were well reproduced
in several independently grown crystals from multiple batches. Both bulk T$%
_{c}$, as measured by 4$\pi \chi $ and resistive T$_{c}$ (as defined in the
text) varied by $\Delta $T$_{c}$ = $\pm $ 1 K. This variation is probably
due to sample degradation induced by variable air exposure. However, the
onset of resistive T$_{c}$ was always above the bulk. Fig. 10 shows
example of superconducting T$_{c}$ and volume fraction for different,
independently grown sample. Figures 11(a) and 11(b) show hysteresis curves
below T$_{c}$ for independently grown crystal from the same batch (sample 2)
as the crystal used in the main text (sample 1). Unlike sample 1 (that was
measured within a day from the moment of its synthesis), sample 2 was
exposed to air for several days. In addition, it was slightly heated when
sealing in quartz tube and kept in the low vacuum (several Torr) for about 3
months. Dominant type-II superconducting MvsH hysteresis is evident whereas
small WFM is superimposed on the main signal. As expected, the magnitude of
WFM signal is about two times larger in sample 2 than in crystal that was
exposed to air for shorter time (Fig. 8(a,b), main text). Based on our synchrotron powder X-ray results taken on samples that were exposed to air for several months, the larger WFM magnitude should also originate from Fe - based compounds (impurities) that form over a course of time (see main text). The thickness of
sample 2 was identical and ab plane was considerably smaller than sample 1,
hence the crystal was more isotropic (a = 0.1535 cm, b = 0.036 cm, c =
0.0036 cm). Larger irreversibility fields are expected when geometric edge
barrier for vortex penetration dwarfs pinning at the inhomogeneities of the
material. If intrinsic, larger WFM signal in sample 2 is expected to
originate from more defects that would cause larger irreversibility field if
the bulk pinning on inhomogeneities is dominant. However, for thin
superconducting strips geometric (specimen shape dependent) barrier is
dominant and larger irreversibility fields are expected for more anisotropic
samples.\cite{Brandt} Both sample 1 and sample 2 are rather thin (c/a =
0.016, c/b = 0.06 for sample 1 and c/a = 0.023, c/b = 0.1 for sample 2),
whereas sample 1 is more anisotropic and is expected to have larger
irreversibility field.

* Present address: Ames Laboratory US DOE and Department of Physics and
Astronomy, Iowa State University, Ames, IA 50011, USA

\ddag\ petrovic@bnl.gov

\end{document}